\documentclass{elsart}
\usepackage{graphicx,amssymb}
\newcommand{\ket}[1]{\vert #1 \rangle}
\newcommand{\bra}[1]{\langle #1 \vert}
\newcommand{\dket}[1]{\vert #1 \rangle\rangle}
\newcommand{\dbra}[1]{\langle\langle #1 \vert}

\begin{document}
\begin{frontmatter}
\title{Optimized teleportation in Gaussian noisy channels}
\author{Stefano Olivares${}^1$, Matteo G. A. Paris${}^2$, and Andrea R. Rossi${}^1$}
\address{${}^1$Dipartimento di Fisica and Unit\`a INFM, Universit\`a degli
Studi di Milano, via Celoria 16, I-20133 Milano, Italia}
\address{${}^2$INFM UdR Pavia, Italia}
\begin{abstract}
We address continuous variable quantum teleportation in Gaussian
quantum noisy channels, either thermal or squeezed-thermal.
We first study the propagation of twin-beam and evaluate a threshold
for its separability. We find that the threshold for purely thermal channels
is always larger than for squeezed-thermal ones. On the other hand, we show
that squeezing the channel improves teleportation of squeezed states and,
in particular, we find the class of squeezed states that are better teleported
in a given noisy channel. Finally, we find regimes where optimized teleportation
of squeezed states improves amplitude-modulated communication in comparison
with direct transmission.
\end{abstract}
\end{frontmatter}
\section{Introduction}\label{s:intro}
In a quantum channel, information is encoded in a set of quantum states, which
are in general nonorthogonal and thus, even in principle, cannot be observed
without disturbance. Therefore, their faithful transmission requires that the
entire communication protocol is carried out by a physical apparatus that
works without knowing or learning anything about the travelling signal. In this
respect, quantum teleportation provides a remarkable mean for indirectly
sending quantum states.
\par
The key ingredient of quantum teleportation is an entangled bipartite state
used to support the quantum communication channel \cite{bennett2}. This allows
the preparation of an arbitrary quantum state at a distant place without
directly transmitting it. In optical implementations of continuous variables
quantum teleportation (CVQT), the entangled source is typically a twin-beam
state of radiation (TWB), whose two modes are shared between the two parties. A
faithful transmission of quantum information through the channel requires a
large input-output fidelity, which in turn is an increasing function of the
amount of entanglement. However, the propagation of a TWB in
noisy channels unavoidably leads to degradation of entanglement, due to
decoherence induced by losses and noise. Indeed, the effect of decoherence on
TWB entanglement and, in turn, on teleportation fidelity, have been addressed
by many authors \cite{wilson,jlee1,jspb,vukics,bowen,progphys}.
Thresholds for separablity of TWB have been established
and teleportation of both classical and nonclassical states
has been explicitly analyzed \cite{banjpa,noise:take}. In particular, in
Ref. \cite{noise:take} it was investigated how much
nonclassicality can be transferred by noisy teleportation in a
zero temperature thermal bath. Moreover, the stability of squeezed states in
a squeezed environment has been recently studied, showing that such
nonclassical states loose their coherence faster than coherent
states even if coupled with nonclassical reservoir \cite{grewal}.
The open question is then if there exist situations where squeezed
states are favoured with respect to coherent ones, especially for
quantum communication purporses.
\par
In this paper we investigate the behavior of a TWB propagating
through a Gaussian noisy channel, either thermal or squeezed-thermal,
and address its performances for applications in quantum communication
\cite{niel:chuang,oliv:paris}.
As we will see, in presence of noise along the channel, teleportation of a
suitable class of squeezed states can be an effective and robust protocol
for amplitude-based communication compared to direct transmission.
\par
Squeezed environments were addressed by many authors for preservation of
the macroscopic quantum coherence. In fact, if squeezed quantum fluctuations 
are added to dissipation, a macroscopic superposition state preserves its
coherence longer than in presence of dissipation alone \cite{sq:ken}.   
Reference \cite{sq:munro} showed that the interference fringes due to a 
superposition of two macroscopically distinct coherent states (``Schr\"odinger's 
cat states'') could be improved by the inclusion of squeezed vacuum
fluctuations.  An interesting physical realization of an environment with
squeezed quantum fluctuations based on quantum-non-demolition-feedback was
proposed in reference \cite{sq:vitali}.  Effective squeezed-bath
interactions were studied in references \cite{sq:clark,sq:clark2}, where
the technique of quantum-reservoir engineering \cite{sq:bath:eng} was
actually used to couple a pair of two-state atoms to an {\em effective}
squeezed reservoir.  
\par
The paper is structured as follows: in sections \ref{s:interaction} and
\ref{s:sep} we describe the evolution of a TWB in a squeezed-thermal bath
and study its separability by means of the partial
transposition criterion; section \ref{s:tele} addresses the TWB coupled
with the non classical environment as a resource for quantum teleportation
of squeezed states; in section \ref{s:comm} we compare the performances of
direct transmission and teleportation. In section \ref{s:conclusions} we
draw some concluding remarks.
\section{Twin beam coupled with a squeezed thermal bath}\label{s:interaction}
The propagation of a TWB interacting with a squeezed-thermal
bath can be modelled as the coupling of each part of the state with
a non-zero temperature squeezed reservoir. The dynamics can be
described by the two-mode Master equation \cite{WM}
\begin{eqnarray}
\frac{d\rho_t}{dt} &=& \{ \Gamma (1+N) L[a]+\Gamma (1+N)
L[b]+\Gamma N L[a^{\dag}] + \Gamma N L[b^{\dag}] \nonumber\\
&\mbox{}& \hspace{1cm}+ \Gamma M \mathcal{M}[a^{\dag}] + \Gamma M^*
\mathcal{M}[a] + \Gamma M \mathcal{M}[b^{\dag}] + \Gamma M^* \mathcal{M}[b]
\}\rho_t \,,
\label{master:squeezed}\;
\end{eqnarray}
where $\rho_t\equiv\rho(t)$ is the system's density matrix at the
time $t$, $\Gamma$ is the damping rate, $N$ and $M$ are the effective
photons number and the squeezing parameter of the bath respectively,
$L[O]$ is the Lindblad superoperator,
$L[O]\rho_t=O\rho_t O^{\dag}-\frac12 O^{\dag}O\rho_t - \frac12
\rho_t O^{\dag} O$, and $\mathcal{M}[O]\rho_t = O\rho_t O -
\frac12 O O \rho_t - \frac12 \rho_t O O $. The terms proportional
to $L[a]$ and $L[b]$ describe the losses, whereas the terms
proportional to $L[a^{\dag}]$ and $L[b^{\dag}]$ describe a linear
phase-insensitive amplification process. Of course, the dynamics
of the two modes are independent on each other.
\par
Thanks to the differential representation of the superoperators in equation
(\ref{master:squeezed}), the corresponding Fokker-Planck equation for the
two-mode Wigner function $W\equiv W(x_1,y_1; x_2,y_2)$ is
\begin{eqnarray}
\partial_t W & = & \frac{\Gamma}{2}\sum_{j=1}^{2}\:
\left(\partial_{x_j} x_j + \partial_{y_j} y_j\right)W +
\frac{\Gamma}{2} \sum_{j=1}^{2}\: \left\{\frac{1}{2}
\Re{\rm e}[M]\left( \partial^{2}_{x_j x_j} - \partial^{2}_{y_j y_j}\right)
\right. \nonumber \\
&\mbox{}& \quad + \left. \Im{\rm m}[M]\: \partial^{2}_{x_j y_j} +
\frac{1}{2}\left(N+\frac{1}{2}\right)\left(\partial^{2}_{x_j x_j}
+ \partial^{2}_{y_j y_j}\right) \right\} W\;,
\end{eqnarray}
which, introducing $\tau = \Gamma t/\gamma$ and $\gamma = (2N+1)^{-1}$,
reduces to the standard form
\begin{eqnarray}
\partial_{\tau} W = \left\{ -\sum_{j=1}^4 \partial_{x_{j}}\:
a_{j}(\underline{x}) + \frac{1}{2} \sum_{i,j=1}^{4}
\partial^{2}_{ x_{i} x_{j}}\: d_{ij} \right\} W\,,
\label{wigner:master:squeezed}
\end{eqnarray}
where, for sake of simplicity, we put $\underline{x} =
(x_{1},y_{1};x_{2},y_{2}) \equiv (x_{1},x_{2};x_{3},x_{4})$. In equation
(\ref{wigner:master:squeezed}) $a_j(\underline{x})$ and
$d_{ij}$ are the matrix elements of the drift and diffusion
matrices $\mathbf{A}(\underline{x})$ and $\mathbf{D}$
respectively, which are given by
\begin{eqnarray}
\mathbf{A}(\underline{x}) = -\frac{\gamma}{2} \,
\underline{x}\,,
\end{eqnarray}
\begin{eqnarray}
\mathbf{D}= \left( \begin{array}{cccc}
\frac{1}{4}+\frac{\gamma}{2} \, \Re{\rm e}[M] & \gamma \, \Im{\rm
m}[M] &
0 & 0 \\
\gamma \, \Im{\rm m}[M] & \frac{1}{4}-\frac{\gamma}{2} \, \Re{\rm
e}[M] &
0 & 0 \\
0 & 0 & \frac{1}{4}+\frac{\gamma}{2} \, \Re{\rm e}[M] & \gamma
 \, \Im{\rm
m}[M] \\
0 & 0 & \gamma \, \Im{\rm m}[M] & \frac{1}{4}-\frac{\gamma}{2} \,
\Re{\rm e}[M]
\end{array} \right)\,.
\end{eqnarray}
Notice that in our case the drift term is linear in $\underline{x}$ and the
diffusion matrix does not depend on $\underline{x}$. We assume $M$ as real and
a TWB as starting state {\em i.e.} $\rho_0\equiv\rho_{\rm TWB}
=|{\rm TWB}\rangle\rangle\langle\langle {\rm TWB} |$, where $|{\rm TWB}
\rangle\rangle= \sqrt{1-x^2} \sum_p\: x^{a^{\dag}a}\: \ket{p}\ket{p}$.
The TWB corresponds to the
Wigner function
\begin{eqnarray}
W_{0}(x_{1},y_{1};x_{2},y_{2}) = \frac{
\exp\left\{
\displaystyle{
- \frac{(x_{1}+x_{2})^{2}}{4\sigma_{+}^{2}}
- \frac{(y_{1}+y_{2})^{2}}{4 \sigma_{-}^{2}}
- \frac{(x_{1}-x_{2})^{2}}{4 \sigma_{-}^{2}}
-\frac{(y_{1}-y_{2})^{2}}{4 \sigma_{+}^{2}}
}
\right\}
}{(2 \pi)^{2}\:
\sigma_{+}^{2}\sigma_{-}^{2}}
\end{eqnarray}
with $\sigma_{\pm}^2 = \frac14 e^{\pm 2 \lambda}$ and $\lambda$, $x=\tanh
\lambda$, being the squeezing parameter of the TWB. Now the solution of the
Fokker-Planck (\ref{wigner:master:squeezed}) is given by \cite{WM}
\begin{eqnarray}
W_{\tau}(x_{1},y_{1},x_{2},y_{2}) = \frac{
\exp\left\{
\displaystyle{
-\frac{(x_{1}+x_{2})^{2}}{4
\Sigma_{1}^{2}}-\frac{(y_{1}+y_{2})^{2}}{4 \Sigma_{2}^{2}}
-\frac{(x_{1}-x_{2})^{2}}{4
\Sigma_{3}^{2}}-\frac{(y_{1}-y_{2})^{2}}{4 \Sigma_{4}^{2}}
}
\right\}
}
{(2 \pi)^2\:
\Sigma_{1}\: \Sigma_{2}\: \Sigma_{3}\: \Sigma_{4}}
\label{wigner:evol:sq}
\end{eqnarray}
where $\Sigma_{j}^2 = \Sigma_{j}^2(\lambda, \Gamma, n_{\rm th}, n_{\rm
s})$, $j=1,2,3,4$, are
\begin{equation}
\begin{array}{lll}
\Sigma_{1}^{2}=\sigma_{+}^{2} e^{-\Gamma t}+D_{+}^{2}(t)\,,
&\quad&
\Sigma_{2}^{2}=\sigma_{-}^{2} e^{-\Gamma t}+D_{-}^{2}(t)\,, \\
\mbox{}\\
\Sigma_{3}^{2}=\sigma_{-}^{2} e^{-\Gamma t}+D_{+}^{2}(t)\,,
&\quad&
\Sigma_{4}^{2}=\sigma_{+}^{2} e^{-\Gamma t}+D_{-}^{2}(t)\,,
\end{array}\label{sigma:1234}
\end{equation}
and
\begin{eqnarray}\label{sq:D:pm}
D_{\pm}^{2}(t) = \frac{1 + 2 N \pm 2 M}{4} \left(1 -e^{-\Gamma
t}\right)
\end{eqnarray}
with $\left| M \right| \leq (2 N + 1)/2$. The latter condition is
already enforced by the positivity condition for the
Fokker-Planck's diffusion coefficient, which requires
\begin{eqnarray} \label{10}
M\leq\sqrt{N(N+1)}\,.
\end{eqnarray}
If we assume the environment as composed by a set of oscillators
excited in a squeezed-thermal state of the form $\nu= S
(r)\rho_{\rm th} S^\dag (r)$, with $S(r)=\exp\{\frac12 r [a^{\dag
2}-a^2]\}$ and $\rho_{\rm th}=(1+n_{\rm th})^{-1} [n_{\rm
th}/(1+n_{\rm th})]^{a^\dag a}$, then we can rewrite the
parameters $N$ and  $M$ in terms of the squeezing and thermal
number of photons $n_{\rm s}=\sinh^2 r$ and $n_{\rm th}$
respectively. Then we get \cite{grewal}
\begin{eqnarray}
M &=&  \left(1 + 2\,n_{\rm th}\right)\sqrt{n_{\rm s}(1 + n_{\rm s})}\,,
\label{phys:par:M} \\
N &=& n_{\rm th} + n_{\rm s}(1 + 2\,n_{\rm th}) \label{phys:par:N}
\:.
\end{eqnarray}
Using this parametrization, the  condition (\ref{10}) is automatically satisfied.
\section{Separability}\label{s:sep}
A quantum state of a bipartite system is {\em separable} if its
density operator can be written as $\varrho=\sum_k p_k \sigma_k
\otimes \tau_k$, where $\{p_k\}$ is a probability distribution and
$\tau$'s and $\sigma$'s are single-system density matrices. If a
state is separable the correlations between the two systems are of
purely classical origin. A quantum state which is not separable
contains quantum correlations {\em i.e.} it is entangled. A
necessary condition for separability is the positivity of the
density matrix $\varrho^T$, obtained by partial transposition of
the original density matrix (PPT condition) \cite{peres}. In
general PPT has been proved to be only a necessary condition for
separability; however, for some specific sets of states, PPT is
also a sufficient condition. These include states of $2\times 2$
and $2 \times 3$ dimensional Hilbert spaces \cite{2x} and Gaussian
states (states with a Gaussian Wigner function) of a bipartite
continuous variable system, {\em e.g.} the states of a two-mode
radiation field \cite{geza,simon}.  Our analysis is based on these
results. In fact, the Wigner function of a twin-beam produced by a
parametric source is Gaussian and the evolution inside active
fibers preserves such character. Therefore, we are able to
characterize the entanglement at any time and find conditions on
the fiber's parameters to preserve it after a given fiber length.
The density matrix's PPT property can be rephrased as a condition
on the covariance matrix of the two modes Wigner function
$W(x_1,y_1;x_2,y_2)$. We have that a state is separable iff
\begin{eqnarray} \label{15}
\mathbf{V}+\frac{i}{4} \mathbf{\Omega}\geq0
\end{eqnarray}
where
\begin{eqnarray}
\begin{array}{cc}
\mathbf{\Omega} = \left( \begin{array}{cc} \mathbf{J} & \mathbf{0} \\
\mathbf{0} & \mathbf{J}
\end{array} \right) &
\mbox{and} \quad
\mathbf{J} = \left( \begin{array}{cc} 0 & 1 \\
-1 & 0
\end{array} \right)\,.
\end{array}
\end{eqnarray}
and
\begin{eqnarray}
V_{pk} = \langle \Delta\xi_p\: \Delta\xi_k \rangle = \int\!\!
{\rm d}^4\xi\: \: \Delta\xi_p\: \Delta\xi_k\: W(\xi)\,,
\end{eqnarray}
with $\Delta\xi_j = \xi_j - \langle \xi_j \rangle$, and
$\underline{\xi}=\{x_1,y_1,x_2,y_2 \}$.
The explicit expression of the covariance matrix associated to the Wigner
function (\ref{wigner:evol:sq}) is
\begin{eqnarray}
\mathbf{V}= \frac{1}{2} \left(
\begin{array}{cccc}
\Sigma_{1}^{2}+\Sigma_{3}^{2} & 0 & \Sigma_{1}^{2}-\Sigma_{3}^{2} & 0 \\
0 & \Sigma_{2}^{2}+\Sigma_{4}^{2} & 0 & \Sigma_{2}^{2}-\Sigma_{4}^{2} \\
\Sigma_{1}^{2}-\Sigma_{3}^{2} & 0 & \Sigma_{1}^{2}+\Sigma_{3}^{2} & 0 \\
0 & \Sigma_{2}^{2}-\Sigma_{4}^{2} & 0 &
\Sigma_{2}^{2}+\Sigma_{4}^{2}
\end{array} \right)\,,
\end{eqnarray}
and, then, condition (\ref{15}) is satisfied when
\begin{eqnarray} \label{20}
\Sigma_{1}^{2}\:\Sigma_{4}^{2} \geq \frac{1}{16}\,,\quad\quad
\Sigma_{2}^{2}\:\Sigma_{3}^{2} \geq \frac{1}{16}\,.
\end{eqnarray}
Notice that changing the sign of $M$ leaves conditions (\ref{20})
unaltered.
\par
\begin{figure}[h!]
\begin{center}
\includegraphics[width=0.7\textwidth]{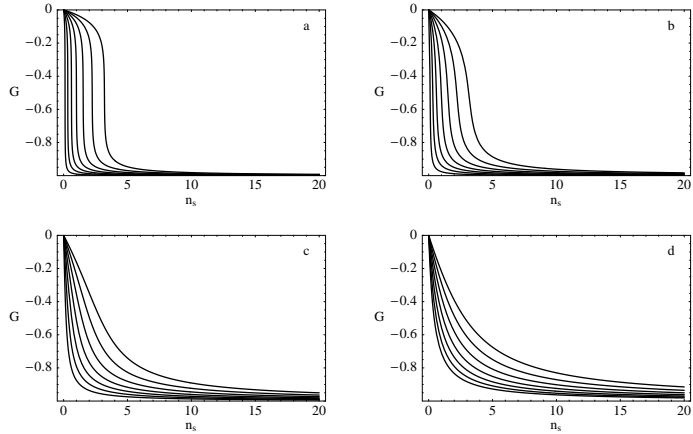}
\caption{Plots of the ratio $G = (t_{\rm s}-t_{0})/t_{0}$ as a function of
the number of squeezed photons $n_{\rm s}$ for different values of the TWB
parameter $\lambda$ and of the number of thermal photons $n_{\rm th}$. The
values of $n_{\rm th}$ are chosen to be: (a) $n_{\rm th} = 10^{-6}$, (b)
$10^{-3}$, (c) $10^{-1}$ and (d) $1$, while the solid lines, from bottom to
top, refer to $\lambda$ varying between 0.1 to 1.0 with steps of
0.15.}\label{f:threshold}
\end{center}
\end{figure}

By solving these inequalities with respect to time, $t$ we find that the two-mode
state becomes separable for $t > t_{\rm s}$, where
the threshold time $t_{\rm s} = t_{\rm s}(\lambda,\Gamma,n_{\rm th},
n_{\rm s})$ is given by
\begin{eqnarray}\label{tau:sep:squeezed}
t_{\rm s} = \frac{1}{\Gamma}\log\left( f + \frac{1}{1+2 n_{\rm th}}
\sqrt{f^2 + \frac{n_{\rm s}(1+n_{\rm s})} {n_{\rm th}(1+n_{\rm
th})}} \right)\,,
\end{eqnarray}
and we defined
\begin{eqnarray}
f \equiv f(\lambda,n_{\rm th}, n_{\rm s}) =\frac{(1+2\,n_{\rm th})\:
\left[ 1+2\,n_{\rm th}-e^{-2\,\lambda}(1+2\,n_{\rm  s}) \right]}
{4\,n_{\rm th}(1+n_{\rm th})}\,.
\end{eqnarray}
As one may expect, $t_{\rm s}$ decreases as $n_{\rm th}$ and
$n_{\rm s}$ increase.  Moreover, in the limit $n_{\rm\rm  s}
\rightarrow 0$, the threshold time (\ref{tau:sep:squeezed})
reduces to the case of a non squeezed bath, in formula \cite{jspb,progphys}
\begin{eqnarray} \label{25}
t_{0} &=&  t_{\rm s}(\lambda,\Gamma,n_{\rm th}, 0)\nonumber\\
&=& \frac{1}{\Gamma} \log \left( 1+\frac{1 - e^{-2 \,
\lambda}}{2\,n_{\rm th}} \right)\,.
\end{eqnarray}
In order to see the effect of a squeezed bath on the entanglement
time we define the function
\begin{eqnarray} \label{26}
G(\lambda, n_{\rm th}, n_{\rm s}) &\equiv& \frac{t_{\rm s}-t_{\rm
0}}{t_0}\,.
\end{eqnarray}
In this way, when $G>0$, the squeezed bath gives a threshold time
longer than the one obtained with $n_{\rm s}=0$, shorter
otherwise. These results are illustrated in figure
\ref{f:threshold}, where we plot equation (\ref{26}) as a function
of $n_{\rm s}$ for different values of $n_{\rm th}$ and $\lambda$.
Since $G$ is always negative, we conclude that coupling a TWB with
a squeezed-thermal bath destroys the correlations between the two
channels faster than the coupling with a non squeezed environment.
\section{Optimized quantum teleportation}\label{s:tele}
In this section we study continuous variable quantum teleportation
(CVQT) assisted by a TWB propagating through a squeezed-thermal environment.
Let us remind the CVQT protocol: the sender and the receiver, say Alice and Bob,
share a two-mode state described by the density matrix $\rho_{12}$, where the
subscripts refer to modes 1 and 2 respectively: mode 1 is sent to Alice, the
other to Bob. The goal of CVQT is teleporting an unknown state $\sigma$,
corresponding to the mode 3, from Alice to Bob. In order to implement the
teleportation, Alice first performs a heterodyne detection on modes 3 and 1,
{\em i.e.} she jointly measures a couple of two-mode quadratures. The POVM
of the measurement is given by
\begin{eqnarray}
\Pi_{13}(z) = \frac{1}{\pi}\:D_1(z) \dket{\mathbb{I}}_{13}
 {}_{31}\dbra{\mathbb{I}} D_{1}^{\dag}(z)\,,\label{het:POVM}\;
\end{eqnarray}
where $\dket{\mathbb{I}}_{13} \equiv \sum_{v}\:\ket{v}_{1}\ket{v}_{3}$, and
$D_1(z) \equiv \exp\{z a^{\dag}-z^* a\}$ is the displacement operator
acting on mode 1.  Each measurement outcome is a complex number $z$, which
is sent to Bob via a classical communication channel, and used by him to
apply a displacement $D(z)$ to mode 2 such to obtain the quantum state
$\rho_{\rm tele}$ which, in an ideal case, coincides with the input signal
$\sigma$ \cite{braun1,furusawa}. The Wigner function of the  heterodyne
POVM is given by \cite{rsp}
\begin{equation}
W[\Pi_{13}(z)](x_1,y_1;x_3,y_3) = \frac{1}{\pi^2}\: \delta\big(
(x_1 - x_3) + x \big)\: \delta\big( (y_1 + y_3) - y \big)\,,
\label{wig:het}\;
\end{equation}
with $z=x+iy$, and since, using Wigner functions, the trace between two
operators can be written as \cite{glauber}
\begin{eqnarray}
{\rm Tr}\:[O_1\:O_2] = \pi\: \int\!\! d^2 w\: W[O_1](w)\: W[O_2](w)\,,
\end{eqnarray}
the heterodyne probability distribution  is given by \cite{oliv}
\begin{eqnarray}
p(z) &=& \pi^3 \int\!\!\!\!\int\!\! dx_1\,dy_1
\int\!\!\!\!\int\!\! dx_2\,dy_2
\int\!\!\!\!\int\!\! dx_3\,dy_3 \: W[\sigma](x_3,y_3)\:\nonumber\\
&\mbox{}& \hspace{1cm}
 \times W[\rho_{12}](x_1,y_1;x_2,y_2)\:
\nonumber\\
&\mbox{}& \hspace{1.5cm} \times W[\Pi_{13}(z)](x_1,y_1;x_3,y_3)\,
W[\mathbb{I}_2](x_2,y_2)\,,
\label{wigner:het:prob}\;
\end{eqnarray}
while the conditional state of mode 2 is
\begin{eqnarray}
W[\rho_2(z)](x_2,y_2) &=& \frac{\pi^2}{p(z)}\:
\int\!\!\!\!\int\!\! dx_1\,dy_1 \int\!\!\!\!\int\!\! dx_3\,dy_3 \:
W[\sigma](x_3,y_3)\:
\nonumber\\
&\mbox{}& \hspace{1cm} \times W[\rho_{12}](x_1,y_1;x_2,y_2)\:
\nonumber\\
&\mbox{}& \hspace{1.5cm} \times W[\Pi_{13}(z)](x_1,y_1;x_3,y_3)\,
W[\mathbb{I}_2](x_2,y_2)\,,
\label{wig:rho:cond}
\end{eqnarray}
where $W[\mathbb{I}_2](x_2,y_2)=\pi^{-1}$. Thanks to equation
(\ref{wig:het}) and after the integration with respect to $x_3$ and $y_3$,
we have
\begin{eqnarray}
W[\rho_2(z)](x_2,y_2) &=& \frac{1}{\pi\,p(z)}\:
\int\!\!\!\!\int\!\! dx_1\,dy_1 \: W[\sigma](x_1+x,-y_1+y)\: \nonumber\\
&\mbox{}& \hspace{2cm} \times
W[\rho_{12}](x_1,y_1;x_2,y_2)\nonumber\\
&=& \frac{1}{\pi\, p(z)}\:
\int\!\!\!\!\int\!\! dx_1\,dy_1 \: W[\sigma](x_1,y_1)\: \nonumber\\
&\mbox{}& \hspace{2cm} \times W[\rho_{12}](x_1-x,-y_1+y;x_2,y_2)\,.
\label{wig:rho:cond:2}
\end{eqnarray}
Now we perform the displacement $D(z)$ on mode 2. Since
$$
W[D(z)\:\rho\:D^{\dag}(z)](x_j,y_j) = W[\rho](x_j-x,y_j-y)\,,
$$
we obtain
\begin{eqnarray}
W[\rho'_2(z)](x_2,y_2) &=& \frac{1}{\pi\,p(z)}\:
\int\!\!\!\!\int\!\! dx_1\,dy_1 \: W[\sigma](x_1,y_1)\: \nonumber\\
&\mbox{}& \hspace{1cm}\times W[\rho_{12}](x_1-x,-y_1+y;x_2-x,y_2-y)\,.
\label{wig:rho:cond:3bis}
\end{eqnarray}
with $\rho'_{2}(z) \equiv D(z)\:\rho_{2}(z)\:D^{\dag}(z)$. The
output state of CVQT is obtained integrating equation
(\ref{wig:rho:cond:3bis}) with respect to all the possible
outcomes of heterodyne detection
\begin{equation}
W[\rho_{\rm tele}](x_2,y_2)=\int\!\! d^2 z\: p(z)\:W[\rho'_2(z)](x_2,y_2)\,.
\label{tele:out:gen}
\end{equation}
Finally, when the shared state is the one given in equation
(\ref{wigner:evol:sq}), equation (\ref{tele:out:gen}) rewrites as follows
\begin{eqnarray}
W[\rho_{\rm tele}](x_2,y_2) &=& \int\!\!\!\!\int\!\! \frac{dx'\:dy'}{4 \pi
\:\Sigma_{2}\:\Sigma_{3}}\: \nonumber\\
&\mbox{}&\hspace{0.5cm} \times
\exp\left\{ - \frac{(x'-x_2)^2}{4 \Sigma_{3}^2}-
\frac{(y'-y_2)^2}{4 \Sigma_{2}^2}\right\}\: W[\sigma](x',y') \\
&=& \int\!\! \frac{d^2 w}{4 \pi \:\Sigma_{2}\:\Sigma_{3}}\:
\exp\left\{ - \frac{(\Re{\rm e}[w])^2}{4\Sigma_{3}^2} -
\frac{(\Im{\rm m}[w])^2}{4 \Sigma_{2}^2}\right\}\:\nonumber\\
&\mbox{}&\hspace{0.5cm} \times
W[D(w)\:\sigma\:D^{\dag}(w)](x_2,y_2)\,, \label{wig:output}
\end{eqnarray}
which shows that the  map $\mathcal{L}$, describing CVQT assisted by
a TWB propagating through a squeezed-thermal environment, is given by
\begin{eqnarray}
\rho_{\rm tele}&\equiv&\mathcal{L}\sigma\nonumber\\
&=& \int\!\! \frac{d^2 w}{4 \pi \:\Sigma_{2}\:\Sigma_{3}}\:
\exp\left\{ - \frac{(\Re{\rm e}[w])^2}{4\Sigma_{3}^2} -
\frac{(\Im{\rm m}[w])^2}{4 \Sigma_{2}^2}\right\}\:\nonumber\\
&\mbox{}&\hspace{0.5cm} \times
D(w)\:\sigma\:D^{\dag}(w)\,,
\label{squeezed:CVQT:map}\;
\end{eqnarray}
{\em i.e.} the teleportation protocol corresponds to a
generalized Gaussian noise. Notice that if $n_{\rm s}\rightarrow 0$,
from equations (\ref{sigma:1234}), (\ref{phys:par:M}) and (\ref{phys:par:N})
one has
\begin{equation}
\Sigma_2^2, \Sigma_3^2 \rightarrow
\sigma_{-}^2\: e^{-\Gamma t} + \frac{1+2 n_{\rm th}}{4}
(1 - e^{-\Gamma t})\,,
\end{equation}
which is the noise due to a thermalized quantum channel \cite{banjpa}.
The map (\ref{squeezed:CVQT:map}) can be extended to the case of a
general Gaussian noise as follows
\begin{equation}
\mathcal{L}_{\rm gen}\sigma =
\int\!\! \frac{d^2 w}{\pi \:\sqrt{{\rm det}[\mathbf{C}]}}\:
\exp\left\{ - \underline{w}\:\mathbf{C}\:\underline{w}^{T}
\right\}\: D(w)\:\sigma\:D^{\dag}(w)\,,
\label{squeezed:CVQT:map:gen}\;
\end{equation}
where $\underline{w}$ is the row vector $\underline{w} = (\Re{\rm e}[w],
\Im{\rm m}[w])$ and $\mathbf{C}$ is the covariance matrix of the noise
\cite{wilson}.
\par
Now, in order to use CVQT as a resource for quantum information processing,
we look for a class of squeezed states which achieves an average
teleportation fidelity greater than the one obtained teleporting coherent
states in the same conditions. The Wigner function of the squeezed state
$\sigma = \ket{\alpha,\zeta} \bra{\alpha,\zeta}$,
$\ket{\alpha,\zeta}=D(\alpha)S(\zeta) |0\rangle$,
is given by (we assume the squeezing parameter $\zeta$ as real)
\begin{eqnarray}
W[\sigma](x_3,y_3) = \frac{2}{\pi}\: \exp\Bigg\{ -
\frac{2\: (x_3 - a)^2}{e^{-2\zeta}} - \frac{2\: (y_3 - b)^2}{e^{2\zeta}} \Bigg\}
\label{w1}\;,
\end{eqnarray}
with $a = \Re{\rm e}[\alpha] $, $b = \Im{\rm m}[\alpha] $. Thanks to
equations (\ref{wig:output}) and (\ref{w1}), we have
\begin{eqnarray}
W[\rho_{\rm tele}](x,y) =
\frac{2\: \exp \left\{
\displaystyle{
- \frac{2\: (x - a)^2}{e^{-2\zeta} + 8\:
\Sigma_{3}^2} - \frac{2\: (y - b)^{2}}{e^{2\zeta} + 8\: \Sigma_{2}^2}
}
\right\} }{\pi \sqrt{(e^{2\zeta} + 8\: \Sigma_{2}^2) (e^{-2\zeta} + 8\:
\Sigma_{3}^2)}}\: \label{tele:squeezed}\;,
\end{eqnarray}
where we suppressed all the subscripts. The average teleportation
fidelity is thus given by
\begin{eqnarray}\nonumber
\overline{F}_{\zeta,{\rm tele}}(\lambda,\Gamma, n_{\rm th}, n_{\rm s})
&\equiv& \pi \int\!\!\!\!\int\!\! {\rm d}x\: {\rm d}y\:
W[\sigma](x,y)\: W[\rho_{\rm out}](x,y)\\
&=& \Big( \sqrt{(e^{2\zeta} + 4\: \Sigma_{2}^2)
(e^{-2\zeta} + 4\: \Sigma_{3}^2)} \Big)^{-1}\,,
\label{squeezed:fid:gen}\end{eqnarray}
which attains its maximum when
\begin{eqnarray}\label{zeta:max}
\zeta = \zeta_{\rm max} \equiv
\frac12 \log \left( \frac{\Sigma_2}{\Sigma_3} \right)\,,
\end{eqnarray}
and, after this maximization, reads as follows
\begin{eqnarray}\label{squeezed:fid:max}
\overline{F}_{\rm tele}(\lambda,\Gamma, n_{\rm th}, n_{\rm s}) =
\frac{1}{1+4\:\Sigma_2\:\Sigma_3}\,.
\end{eqnarray}
\par
For $n_{\rm s} \rightarrow 0$ we have $\Sigma_2 = \Sigma_3$, and
thus then $\zeta_{\rm max}\rightarrow 0$, {\em i.e.} the input
state that maximizes the average fidelity (\ref{squeezed:fid:gen})
reduces to a coherent state. In other words, in a non squeezed
environment the teleportation of coherent states is more effective
than that of squeezed states. Moreover, equation
(\ref{squeezed:fid:max}) shows that meanwhile the TWB becomes
separable, {\em i.e.} $\Sigma_2^2\:\Sigma_3^2 \ge \frac{1}{16}$
(see equations (\ref{20})), one has $\overline{F}_{\rm tele}\leq
0.5$. We remember that when the average fidelity is less than 0.5,
the same results can be achieved using {\em classical} (non
entangled) shared states \cite{braun1,braun2}: in our case, it
could be possible to verify the separability of the shared state
simply studying the fidelity achieved teleporting squeezed states.
Notice that the classical limit $\overline{F}_{\rm tele} = 0.5$,
which was derived in the case of coherent state teleportation
\cite{braun2}, still holds when we wish to teleport a squeezed
state with a fixed squeezing parameter. Finally, the asymptotic
value of $\overline{F}_{\rm tele}$ for $\Gamma t \rightarrow
\infty$ is
\begin{eqnarray}
\overline{F}_{\rm tele}^{(\infty)} = \frac{1}{2\: (1 + n_{\rm th})}\,,
\end{eqnarray}
which does not depend on the number of squeezed photons and is
equal to $0.5$ only if $n_{\rm th} = 0$. This last result is
equivalent to say that in presence of a zero-temperature
enviroment, no matter if it is squeezed or not, the TWB is
non-separable at every time.
\par
\begin{figure}[h!]
\begin{center}
\includegraphics[width=0.7\textwidth]{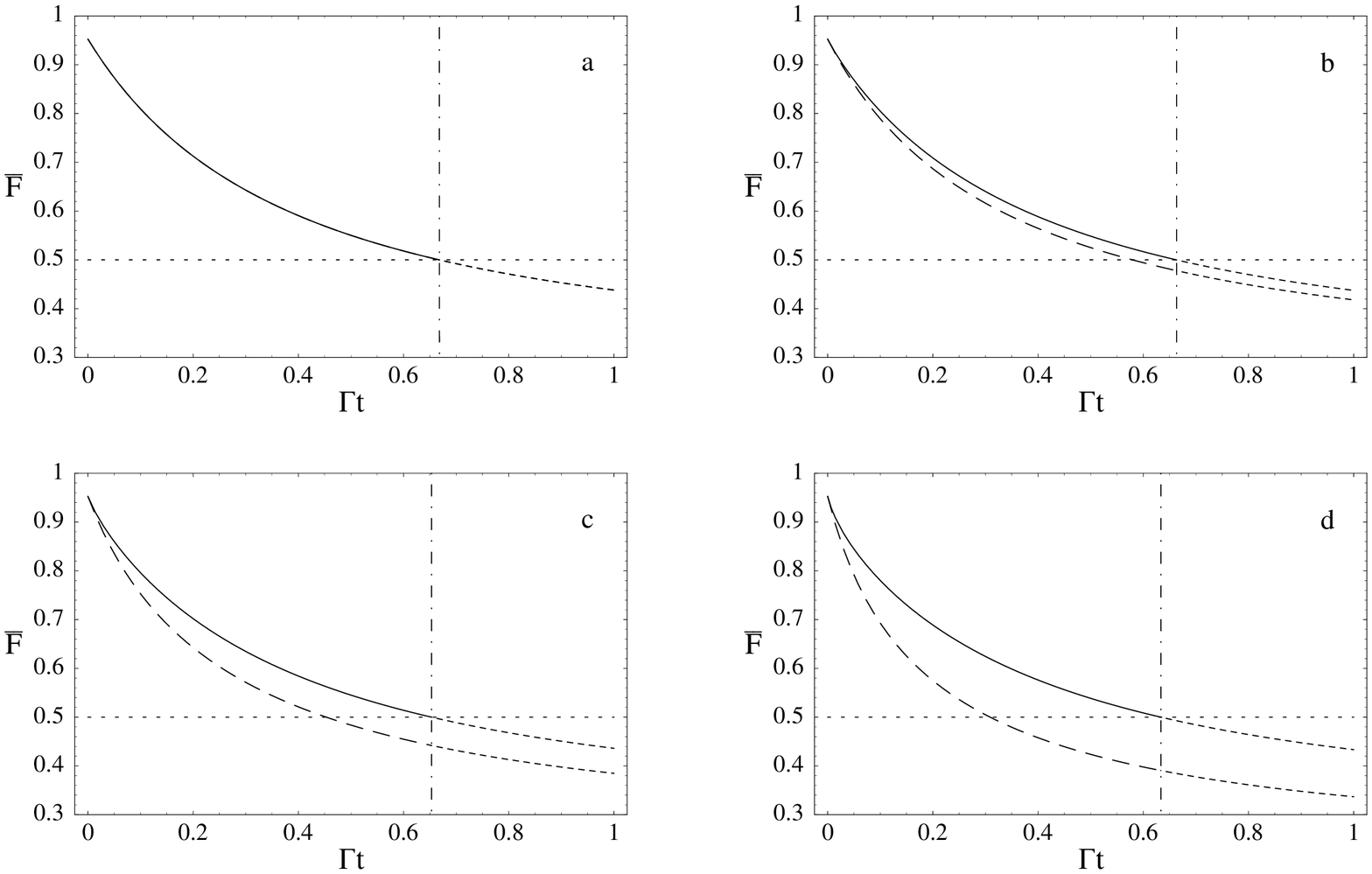}
\caption{Plots of the average teleportation fidelity. The solid and the
dashed lines represent squeezed and coherent state fidelity, respectively,
for different values of the number of squeezed photons $n_{\rm s}$: (a)
$n_{\rm s} = 0$, (b) $0.1$, (c) $0.3$, (d) $0.7$. In all the plots we put
the TWB parameter $\lambda = 1.5$ and number of thermal photons $n_{\rm th}
= 0.5$.  The dot-dashed vertical line indicates the threshold $\Gamma
t_{\rm s}$ for the separability of the shared state: when $\Gamma t >
\Gamma t_{\rm s}$ the state is no more entangled. Notice that, in the case
of squeezed state teleportation, the threshold for the separability
corresponds to $\overline{F} = 0.5$.}\label{f:fidelity}
\end{center}
\end{figure}

In figure \ref{f:fidelity} we plot $\overline{F}_{\rm tele}$ as a function
of $\Gamma t$ for different values of $\lambda$, $n_{\rm th}$ and $n_{\rm
s}$.  As $n_{\rm s}$ increases, the non classicity of the thermal bath
starts to affect the teleportation fidelity and we observe that the
best results are obtained when the state to be teleported is the squeezed
state that maximizes (\ref{squeezed:fid:gen}). Furthermore the difference
between the two fidelities increases as $n_{\rm s}$ increases.  Notice that
there is an interval of values for $\Gamma t$ such that the coherent state
teleportation fidelity is less than the classical limit $0.5$, although
the shared state is still entangled.
\section{Teleportation vs direct transmission}\label{s:comm}
\begin{figure}[h!]
\begin{center}
\includegraphics[width=0.7\textwidth]{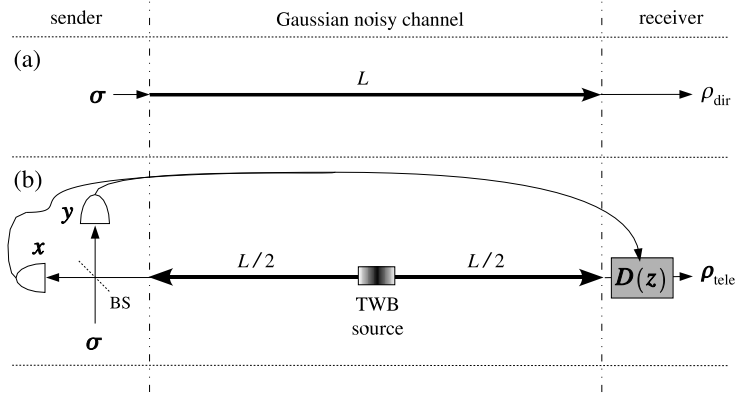}
\caption{ Direct and teleportation-assisted transmission. (a) In
direct transmission, the sender directly sends state $\sigma$
through the Gaussian noisy channel: the state arriving at the
receiver is $\rho_{\rm dir}$. (b) In teleportation-assisted
transmission, the sender mixes at the balanced beam splitter BS
the state $\sigma$ to be transmitted with one of the two mode of
the shared state, arriving from the Gaussian noisy channel, and
then he measures the quadrature $x$ and $y$, respectively, of the
output modes. This result is classically communicated to the
receiver, which applies a displacement $D(z)$, $z=x+iy$, to the
output state, obtaining $\rho_{\rm tele}$ (see section
\ref{s:tele} for details). Notice that the length of the direct
transmission line is twice the effective length of the
teleportation-assisted transmission one.}\label{f:scheme}
\end{center}
\end{figure}

This section is devoted to investigate whether the results obtained in the
previous sections can be used to improve quantum communication using
non classical states. We suppose to have a communication protocol
where information is encoded onto the field amplitude of a set of squeezed
states of the  form $\ket{\alpha,\zeta}$ with fixed squeezing parameter.
In figure \ref{f:scheme} we show a schematic diagram for direct
and teleportation-assisted communication. As one
can see from the figure, direct transmission line's length $L$ is
twice the effective length of the teleportation-assisted scheme:
this is due to the fact that the two modes of the shared state are
chosen to be propagating in opposite directions.
\par
When we directly send the squeezed state (\ref{w1}) through a squeezed
noisy quantum channel, the state arriving at the receiver is
\begin{eqnarray}
W[\rho_{\rm dir}](x,y) =
\frac{2\: \exp \left\{
\displaystyle{
- \frac{2\: (x - a\: e^{-\Gamma t'/2})^2}{e^{-2\zeta- \Gamma t'} + 4
D_{+}^2} - \frac{2\: (y - b\: e^{-\Gamma t'/2})^{2}}{e^{2\zeta
- \Gamma t'} + 4 D_{-}^2}}\right\}}
{\pi \sqrt{(e^{-2\zeta - \Gamma t'} + 4 D_{+}^2) (e^{2\zeta -
\Gamma t'} + 4 D_{-}^2)}}\: \label{evol:squeezed}\;,
\end{eqnarray}
with $D_{\pm}^2$, evaluated at time $t'$, given in equation
(\ref{sq:D:pm}) and time $t'$ is twice the time $t$ implicitly
appearing in equation (\ref{tele:squeezed}), because of the
previously explained choice. Equation (\ref{evol:squeezed}) is
the Wigner function of the state $\rho_{\rm dir}$, solution of the
single-mode Master equation
\begin{eqnarray}
\frac{{d}\rho_t}{{d}t} = \{ \Gamma (1+N) L[a]+ \Gamma N L[a^{\dag}] +
\Gamma M \mathcal{M}[a^{\dag}] + \Gamma M^* \mathcal{M}[a] \}\rho_t \,,
\label{sm:master:squeezed}\;
\end{eqnarray}
where $\Gamma$, $N$, $M$ and the superoperators $L[O]$ and
$\mathcal{M}[O]$ have the same meaning as in equation
(\ref{master:squeezed}).  As in case of quantum teleportation,
we can define the direct transmission fidelity (see equation
(\ref{squeezed:fid:gen})), obtaining
\begin{eqnarray}
F_{\zeta,\alpha,{\rm dir}}(\Gamma, n_{\rm th}, n_{\rm s}) =
\frac{\exp \left\{
\displaystyle{
- \frac{a^2 (1 - e^{-\Gamma t'/2})^2}
{2\:\Sigma_{a}^2(t')} -
\frac{b^2 (1 + e^{-\Gamma t'/2})^{2}(t')}
{2\:\Sigma_{b}^2(t')}}\right\}}
{2\sqrt{
\Sigma_{a}^2(t')\:\Sigma_{b}^2(t')}}\: \label{squeezed:dir:fid:gen}\;.
\end{eqnarray}
where
\begin{eqnarray}
\Sigma_{a}^2(t') &=& \mbox{$\frac14$} e^{-2\zeta}(1 + e^{-\Gamma t'})
+ D_{+}^2(t')\;,\\
\Sigma_{b}^2(t') &=& \mbox{$\frac14$} e^{2\zeta}(1 + e^{-\Gamma t'})
+ D_{-}^2(t')\,.
\end{eqnarray}
Since $F_{\zeta,\alpha,{\rm dir}}$ depends on the amplitude $\alpha = a +i
b$ of the state to be transmitted, in order to evaluate the average fidelity
here we assume that the trasmitter sends squeezed states with fixed
squeezed parameter and with amplitudes distributed according to the Gaussian
\begin{eqnarray}
\mathcal{P}(\alpha) = \frac{1}{2 \pi \Delta^2}
\exp\Bigg\{-\frac{|\alpha|^2}{2\Delta^2}\Bigg\}
\label{gaussian:comm:distrib}\,.
\end{eqnarray}
The average direct transmission fidelity reads as follows
\begin{eqnarray}
\overline{F}_{\zeta,{\rm dir}} &=&
\int\!\! d^2\alpha\: \mathcal{P}(\alpha)\:F_{\zeta,\alpha,{\rm
dir}}(\Gamma, n_{\rm th}, n_{\rm s})\\
&=& \frac12 \Bigg\{\sqrt{
\big[ (1 - e^{-\Gamma t'/2})\Delta^2 + \Sigma_{a}^2(t')\big]
\big[ (1 - e^{-\Gamma t'/2})\Delta^2 + \Sigma_{b}^2(t')\big]
}\Bigg\}^{-1}
\label{direct:fid:gen}\;,
\end{eqnarray}
which, for $\Gamma t \rightarrow \infty$ and using equation (\ref{zeta:max}),
reduces to
\begin{eqnarray}
\overline{F}_{{\rm dir}}^{(\infty)} =
\frac12 \Big(\sqrt{g_{+}(n_{\rm th}, n_{\rm s})\:
g_{-}(n_{\rm th}, n_{\rm s})}\Big)^{-1} \,,
\end{eqnarray}
with
\begin{eqnarray}
g_{\pm}(n_{\rm th}, n_{\rm s}) &=&
\frac14 \Bigg[
1 + \sqrt{1 + 8\: n_{\rm s}(1 + n_{\rm s}) \pm 4\:(1 + 2\:n_{\rm s})
\sqrt{n_{\rm s}(1 + n_{\rm s})}}
\nonumber\\
&\mbox{}&\hspace{0.5cm}
+ 2 \Big(
n_{\rm s} + n_{\rm th} + 2\: n_{\rm s} n_{\rm th} \pm (1 + 2\: n_{\rm th})
\sqrt{n_{\rm s }(1 + n_{\rm s})}
\Big)
\Bigg] + \Delta^{2}\,.\nonumber
\end{eqnarray}
\par
Teleportation is a good resource for quantum communication in noisy channel
when $\overline{F}_{\zeta,{\rm tele}}\geq\overline{F}_{\zeta,{\rm dir}}$,
which gives a threshold $\Delta_{\rm th}^2$ on the width $\Delta^2$ of the
distribution (\ref{gaussian:comm:distrib})
\begin{eqnarray}
\Delta_{\rm th}^2(\lambda,\Gamma, n_{\rm th}, n_{\rm s}) &=&
\frac{1}{2\:(1 - e^{-\Gamma t})^2}
\bigg\{
-\big[\Sigma_{a}^2(2t) + \Sigma_{b}^2(2t)\big]\nonumber\\
&\mbox{}& \hspace{0.5cm}
+ \sqrt{\big[\Sigma_{a}^2(2t) - \Sigma_{b}^2(2t)\big]^2 +
\big( \overline{F}_{\zeta,{\rm tele}} \big)^{-2}}
\bigg\}
\label{Delta:threshold}\;,
\end{eqnarray}
where $\overline{F}_{\zeta,{\rm tele}}$ of Eq. (\ref{squeezed:fid:gen})
is evaluated at time $t$ and,
then, $t' = 2t$.
\par
\begin{figure}[h!]
\begin{center}
\includegraphics[width=0.7\textwidth]{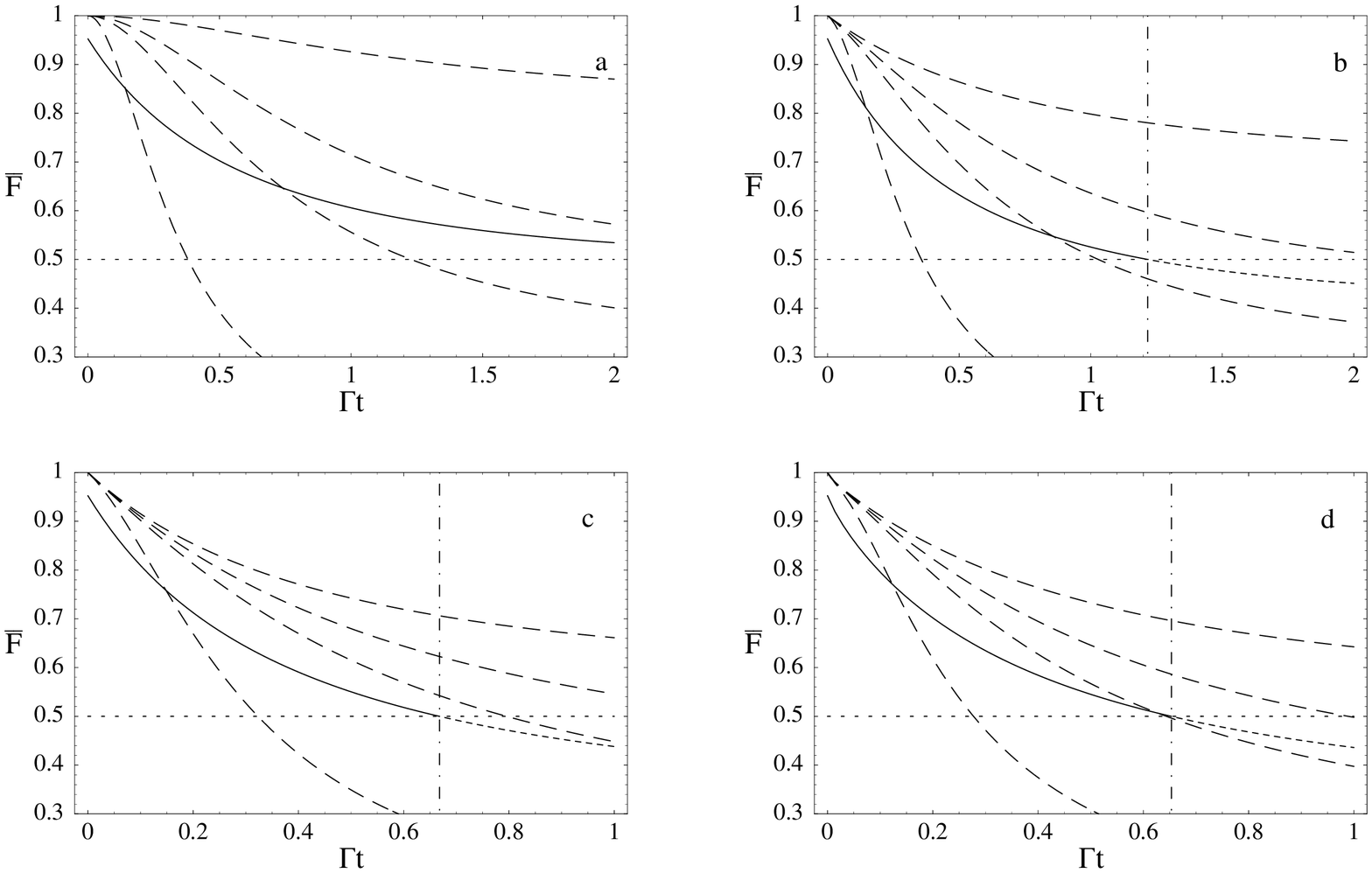}
\caption{Plots of average teleportation (equation (\ref{squeezed:fid:gen}))
and direct communication (equation (\ref{direct:fid:gen})) fidelity as
functions of $\Gamma t$ for different values of $n_{\rm th}$ and $n_{\rm
s}$: (a) $n_{\rm th} = n_{\rm s} = 0$; (b) $n_{\rm th} = 0.3$ and $n_{\rm
s} = 0$; (c) $n_{\rm th} = 0.5$ and $n_{\rm s} = 0$: (d) $n_{\rm th} = 0.5$
and $n_{\rm s} = 0.3$. In all the plots the solid line referes to
$\overline{F}_{\rm tele}$ with $\lambda = 1.5$, whereas the dashed lines
are $\overline{F}_{\rm dir}$ with (from top to bottom) $\Delta^2 = 0.1,
0.5, 1, 5 $. The squeezing parameter is chosen to be $\zeta = \zeta_{\rm
max}$, which maximizes teleportation fidelity. Notice that direct
transmission fidelity is evaluated in a time $t$ equal twice the time of
teleportation (see the scheme in figure \ref{f:scheme}).  The dot-dashed
vertical line indicates the threshold $\Gamma t_{\rm s}$ for the
separability of the shared state used in teleportation.}\label{f:comm}
\end{center}
\end{figure}

In figure \ref{f:comm} we plot $\overline{F}_{\zeta,{\rm tele}}$ and
$\overline{F}_{\zeta,{\rm dir}}$ with $\zeta = \zeta_{\rm max}$ for
different values of the other parameters. We see that teleportation is an
effective and robust resource for communication as the channel becomes more
noisy and $\Delta^2$ larger. Moreover, when $n_{\rm th}, n_{\rm s}
\rightarrow 0$, one obtains the following finite value for the threshold
\begin{eqnarray}
\Delta_{\rm th}^2(\lambda,\Gamma,0,0) = \frac{e^{\Gamma t} - 1 + e^{-2
\lambda}}{2\:e^{-\Gamma t}(1 - e^{-\Gamma t})^2}\,,
\end{eqnarray}
{\em i.e.} teleportation assisted communication can be more effective
than direct transmission even for pure dissipation at zero temperature.
\section{Conclusions}\label{s:conclusions}
In this work we have studied the propagation of a TWB through a
Gaussian quantum noisy channel, either thermal or squeezed-thermal, and have
evaluated the threshold time after which the state becomes separable.
Moreover, we have explicitly found the completely positive map for the
teleportated state using the Wigner formalism.
\par
We have found that the threshold for a squeezed environment is always
shorter than for a purely thermal one. On the other hand, we have shown
that squeezing the channel is a useful resource when entanglement is used
for teleportation of squeezed states. In particular, we have found the class
of squeezed states which optimize teleportation fidelity. The squeezing
parameter of such states depends on the channel parameters themselves.
In these conditions, the teleportation fidelity is always larger than the one
achieved by teleporting coherent states. Moreover, there are no regions of
useless entanglement, {\em i.e.} the fidelity approaches the classical limit
$\overline{F}=0.5$ when the TWB becomes separable.
\par
Finally, we have found regimes where the optimized teleportation
of squeezed states can be used to improve the transmission of
amplitude-modulated signals through a squeezed-thermal noisy channel.
The transmission performances have been investigated by means of input-output
fidelity, comparing the direct transmission with the teleportation one.
Actually, decoherence mechanisms are different between these two channels: in the
teleportation channel the fidelity is reduced due to the interaction of the TWB
with the squeezed-thermal bath; in direct transmission the signal is
directly coupled with the non-classical environment and, then, fidelity is
affected by the degradation of the signal itself. The performance of CVQT
as a quantum communication channel in nonclassical environment obviously
depends on the parameters of the channel itself, but our analysis has shown
that if the signal is drawn from the class of squeezed states that optimize
teleportation fidelity, and the probability distribution of the transmitted
state amplitudes is wide enough, then teleportation is more effective and
robust as the environment becomes more noisy.

\end{document}